\documentclass[twocolumn]{aa} 
\usepackage{graphicx}
\usepackage{txfonts}
\usepackage{xcolor}
\usepackage{comment}

\usepackage[colorlinks, citecolor=blue]{hyperref}
\begin{document}

   \title{CO(1--0) imaging reveals 10-kiloparsec molecular gas reservoirs around star-forming galaxies at high redshift}

   \author{Matus Rybak \inst{1,2,3},
        J. T. Jansen \inst{1},
         M. Frias Castillo \inst{1},
         J. A. Hodge \inst{1},
         P. P. van der Werf \inst{1},
         I. Smail\inst{4},
         G. Calistro Rivera \inst{5,6},
         S. Chapman \inst{7},
         C.-C. Chen \inst{8},
         E. da Cunha \inst{9,10},
         H. Dannerbauer\inst{11},
         E. F. Jim\'enez-Andrade \inst{12},
         C. Lagos\inst{9,10,13},
         C.-L. Liao \inst{1,8},
         E. J. Murphy\inst{14},
         D. Scott\inst{15},
         A. M. Swinbank\inst{4},
         F. Walter\inst{16}}
          
    \institute{
        Leiden Observatory, Leiden University, P.O. Box 9513, 2300 RA Leiden, The Netherlands \\ \email{mrybak@strw.leidenuniv.nl}
        \and Faculty of Electrical Engineering, Mathematics and Computer Science, Delft University of Technology, Mekelweg 4, 2628 CD Delft, The Netherlands
        \and SRON -- Netherlands Institute for Space Research, Niels Bohrweg~4, 2333 CA Leiden, The Netherlands
        \and Centre for Extragalactic Astronomy, Department of Physics, Durham University, South Road, Durham DH1 3LE, UK
        \and German Aerospace Center (DLR), Institute of Communications and Navigation, Wessling, Germany
        \and European Southern Observatory (ESO), Karl-Schwarzschild-Stra\ss e 2, 85748 Garching bei M\"unchen, Germany
        \and Department of Physics and Atmospheric Science, Dalhousie University, Halifax, NS, B3H 4R2, Canada
        \and Academia Sinica Institute of Astronomy and Astrophysics (ASIAA), No. 1, Section 4, Roosevelt Road, Taipei 106216, Taiwan
        \and International Centre for Radio Astronomy Research, University of Western Australia, 7 Fairway, Crawley, Perth, Australia
        \and ARC Centre of Excellence for All Sky Astrophysics in 3 Dimensions (ASTRO 3D), Australia
        \and Cosmic Dawn Center (DAWN), Denmark
        \and Instituto de Astrof\'isica de Canarias, V\'ia L\'actea, 39020 La Laguna (Tenerife), Spain
        \and Instituto de Radioastronom\'ia y Astrof\'isica, Universidad Nacional Aut\'onoma de M\'exico, Antigua Carretera a P\'atzcuaro \# 8701, Ex-Hda. San Jos\'e de la Huerta, Morelia, Michoac\'an, M\'exico C.P. 58089
        \and National Radio Astronomy Observatory, Charlottesville, VA, USA 
        \and Department of Physics and Astronomy, University of British Columbia, 6224 Agricultural Road, Vancouver, BC V6T 1Z1, Canada
        \and Max–Planck Institut f\"ur Astronomie, Königstuhl 17, 69117 Heidelberg, Germany
    }
   \date{Received 12 November 2024; Revised 27 May 2025 }

  \abstract
  {
  Massive, intensely star-forming galaxies at high redshift require a supply of molecular gas from their gas reservoirs, replenished by infall from the surrounding circumgalactic medium, to sustain their immense star-formation rates. However, our knowledge of the extent and morphology of their cold-gas reservoirs is still in its infancy. 
  We present the results of stacking 80 hours of JVLA observations of CO(1--0) emission -- which traces the cold molecular gas -- in nineteen $z=2.0-4.5$ dusty, star-forming galaxies from the AS2VLA survey. The visibility-plane stack reveals extended emission with a half-light radius of $3.8\pm0.5$~kpc, 2--3$\times$ more extended than the dust-obscured star formation and $1.4\pm0.2\times$ more extended than the stellar emission revealed by JWST. Stacking the [\ion{C}{i}](1--0) observations for ten galaxies from our parent sample yields a half-light radius $\leq$2.6~kpc, marginally smaller than CO(1--0). The CO(1--0) size is also comparable to the [\ion{C}{ii}] halos detected around high-redshift star-forming galaxies, {suggesting these arise from molecular gas}. {Photo}-dissociation region modelling indicates that the extended CO(1--0) emission arises from clumpy, dense clouds rather than smooth, diffuse gas.
  Our results show that the bulk (up to 80\%) of molecular gas resides outside the star-forming region; with only a small part directly contributing to their current star formation.   }


\titlerunning{Extended CO(1--0) emission around high-redshift galaxies}
\authorrunning{M. Rybak, J. Jansen, et al.}
   \maketitle
%

\section{Introduction}
\label{sec:introduction}

The majority of star formation beyond redshift $z\approx1$ took place in massive, dusty star-forming galaxies (DSFGs\footnote{Also know as sub-millimetre galaxies (SMGs).}) with star-formation rates (SFRs) of a few 100 -- 1000~M$_\odot$ yr$^{-1}$ (e.g., \citealt{Magnelli2011,Dudzeviciute2020, Zavala2021}). While some DSFGs are starbursts triggered by major mergers, many appear to be relatively isolated \citep{Gillman2024, Hodge2024}. The immense SFRs of these isolated galaxies need to be fed by a continuous supply of cold gas from their extended gas reservoirs and the surrounding circumgalactic medium (CGM, e.g., \citealt{Tumlinson2017, FaucherOh2023}) or by minor mergers with gas-rich satellites (e.g., \citealt{narayanan2015}). 

In local galaxies, cold, molecular gas is primarily traced by the CO(1--0) rotational line, which has an excitation temperature of just 5.5~K and critical density of $2\times10^3$~cm$^{-3}$. At high redshift, studies of molecular gas have mostly focused on higher-excitation CO rotational emission lines (see, e.g., \citealt{hodge2020, tacconi2020,walter2020, birkin2021}). However, the mid/high-excitation CO emission lines arise from the denser, warmer gas, and are expected to be much more compact than the ground-state CO(1--0) transition (e.g., \citealt{Leroy2009, ivison2011, emonts2016}), making them unsuitable for tracing cold, diffuse gas. 

In contrast to the higher-excitation CO lines, studies of the ground-state CO(1--0) line {in high-redshift} galaxies have been scarce and largely focused on the few brightest sources. This is because of the low intrinsic brightness of the CO(1--0) line, {which requires} a large collecting area and long integrating times to achieve the necessary signal-to-noise ratio (S/N).  

The CO(1--0) emission in a number of high-redshift galaxies has been observed by the Green Bank Telescope (e.g., \citealt{hainline2006, swinbank2010, frayer2011, frayer2018, harris2012, harrington2021}). However, these spatially unresolved observations do not provide any information about the source size. Early Karl. G. Jansky Very Large Array (JVLA) studies of CO(1--0) emission in bright high-redshift DSFGs (e.g., \citealt{ivison2010, ivison2011, riechers2011, thompson2012}) reported half-light radii ranging from a few kpc to $\approx$25~kpc (J02399-0136, \citealt{thompson2012, frayer2018}), often showing multiple CO(1--0) components. However, some of these results have been challenged by deeper observations: for example, \citet{friascastillo2022} showed that the supposedly extended CO(1--0) reservoir with high-velocity wings in the $z\approx3.4$ galaxy J13120+4242 \citep{riechers2011b} is likely an artefact of low signal-to-noise of the original data. To summarise, the sizes of CO(1--0) reservoirs reported by these early studies might be biased towards the brightest (i.e., most extreme) sources, and suffer from low S/N.

More recently, apparently very extended molecular gas reservoirs have been reported in high-density environments of early protoclusters, such as the Spiderweb ($z\approx2.2$, \citealt{emonts2016}) which shows a CO(1--0) reservoirs with a half-light radius of $R_{1/2}\approx30$~kpc CO(1--0), the $\approx40$-kpc CO(4--3) emission around the $z\approx3.4$ star-forming galaxy from \citet{Ginolfi2017}, or the claimed super-extended, $\approx$200~kpc CO(3--2) emission around a $z=2.2$ quasar cid\_346 \citep{Cicone2021}, {which was however not confirmed with deeper data} \citep{Jones2023}).

Further evidence for extended gas around high-redshift DSFGs comes from Atacama Large (Sub)Millimetre Array (ALMA) observations of the [\ion{C}{ii}] 158-$\mu$m line. Extended [\ion{C}{ii}] ``halos'' have been observed around high-redshift DSFGs (e.g., \citealt{gullberg2018, rybak2019, rybak2020}) as well as {around} less massive star-forming galaxies (e.g., \citealt{Fujimoto2019, Fujimoto2020, ginolfi2020, Ikeda2025}), sometimes on scales as large as 50~kpc. 

The nature of the [\ion{C}{ii}] halos is still contested: these could be either low-mass satellite galaxies, outflows from the central galaxy (e.g., \citealt{pizzati2020, Pizzati2023}) or incoming streams of cold, molecular gas. In particular, interpreting the [\ion{C}{ii}] observation is difficult: the [\ion{C}{ii}] 158-$\mu$m emission arises from all gas phases (molecular -- H$_2$, atomic -- H$^0$, and ionised -- H$^+$), rather than just the molecular phase. 

The ground-state CO(1--0) emission line remains the most robust probe of diffuse molecular gas. In the recent years, there have been systematic surveys of CO(1--0) in high-redshift DSFGs using JVLA \citep{FriasCastillo2023, Stanley2023}. These have yielded tentative evidence for extended CO(1--0) emission: \citet{FriasCastillo2023} found that the CO(1--0) extracted from moment-0 maps increases with the aperture size out to radii of $\approx$5$"$ (40~kpc). Similarly, using $\approx$1$"$-resolution JVLA imaging of 14 galaxies from the z-GAL survey \citep{Cox2023}, \citet{Stanley2023} found CO(1--0) reservoirs of up to 25~kpc across. However, {as about 50\% of the \citet{Stanley2023} sources are gravitationally lensed}, the interpretation of the measured source sizes is not straightforward.

In this Paper, we report results of stacking 80 hours of CO(1--0) observations from the recently completed AS2VLA\footnote{AS2VLA = ALMA, SCUBA-2, VLA.} survey \citep{FriasCastillo2023}. 
{Specifically, using the JVLA in the most compact D-array and aiming at high surface-brightness sensitivity, provides the ideal set-up to study potentially very extended, low surface-brightness cold gas reservoirs around high-redshift DSFGs.}


{This Paper is structured as follows: in Section~\ref{sec:observations}, we describe our sample selection, details of observations, and the visibility-plane analysis and stacking procedure. We then look for potential cold-gas outflows (Section~\ref{subsec:spectral_stack}), present the ancillary [\ion{C}{i}](1--0) data, and use PDR modelling to infer the thermodynamics of extended gas reservoirs (\ref{subsec:pdr}). Finally, we compare our inferred CO(1--0) sizes to observations of dust, [\ion{C}{ii}], and stellar emission (Section \ref{subsec:sizes} and \ref{subsec:m_mol}) and predictions from hydrodynamical simulations (\ref{subsec:sims}).}

Throughout this paper we assume a standard flat $\Lambda$CDM cosmology with $H_0$ = 67.8 km s$^{-1}$ Mpc$^{-1}$, $\Omega_\mathrm{M}$ = 0.310 and $\Omega_\Lambda$ = 0.690 \citep[][]{planck2016}.

\begin{figure*}[]
    \centering
     \includegraphics[height=5cm]{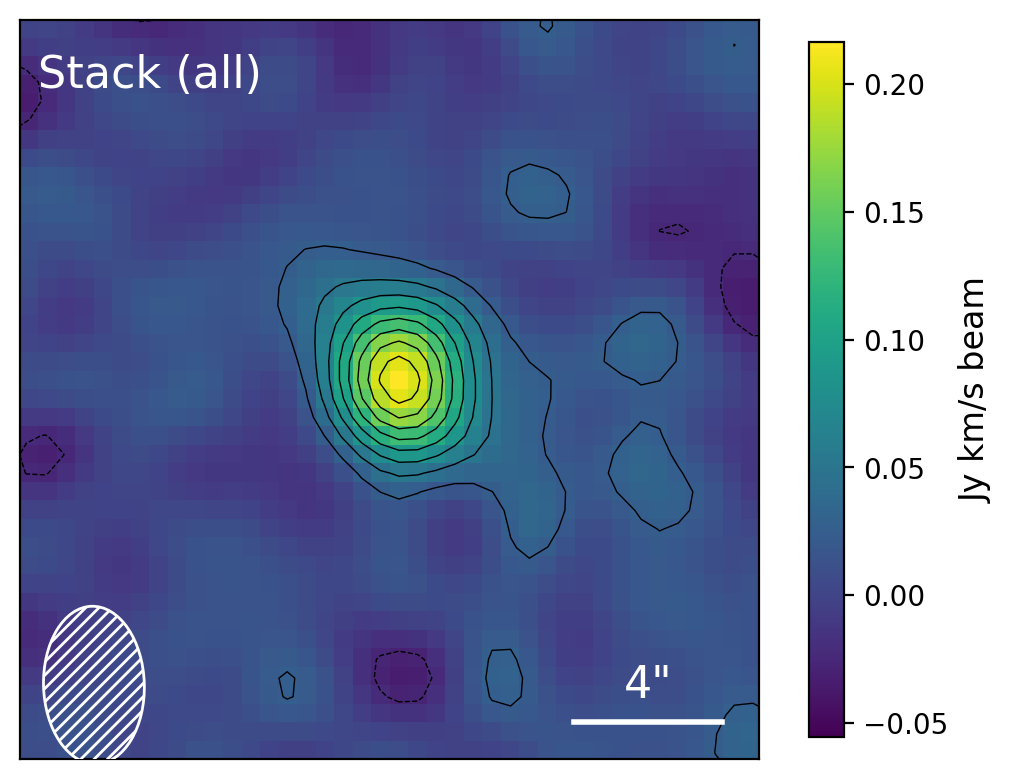}
     \includegraphics[height=5cm]{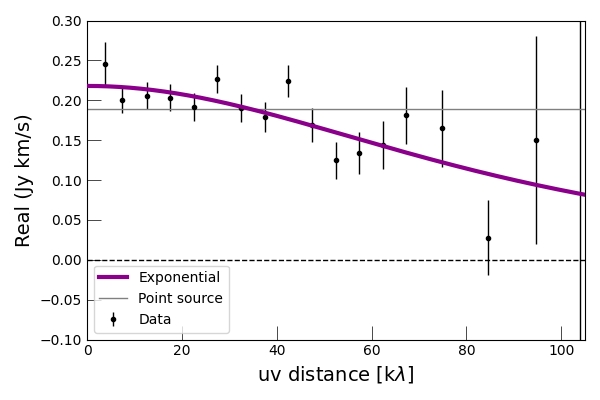}\\
        
    \caption{\textit{Left}: Stacked CO(1--0) image-plane for 19 sources with robust CO(1--0) detections. 
    For the image-plane stack, the contours are drawn at $\pm(2,4,6...)\sigma$, {where 1~$\sigma$=0.014~Jy km/s)} the white ellipses indicate the mean FWHM beam size (4.24''$\times$2.73'', PA = 1.7~deg). \textit{Right}: Visibility-plane stack. The $uv$-plane data are integrated over $\pm$1~FWHM velocity range and radially binned with a step of 5~k$\lambda$. For CO(1--0), the exponential model with $R_\mathrm{1/2}$=$0.49\pm0.07$" (3.8$\pm$0.5~kpc at our median redshift $z=3.1$) is strongly preferred by the available data.}
    \label{fig:stack_image}
\end{figure*}

\section{Observations and imaging}
\label{sec:observations}

\subsection{Target sample}
\label{subsec:target_sample}
Our initial sample consists of 30 sub-millimetre selected, star-forming galaxies at $z=2.2-4.4$ from the AS2VLA survey \citep[Paper~I]{FriasCastillo2023}. The sample is selected from ALMA continuum detections of source taken from $\approx$4~deg$^2$ of SCUBA-2 850$\mu$m maps from the S2CLS \citep{geach2017} and S2COSMOS \citep{simpson2019} surveys. Individual sources were pin-pointed by high-resolution ALMA and SMA imaging \citep{stach2018, hill2018, simpson2020}; the 870-$\mu$m fluxes range between 7.5 and 18.3~mJy. All sources have spectroscopic redshifts based on mid-$J$ CO lines and [\ion{C}{i}](1--0) (\citealt{birkin2021, Chen2022, AS2COSPEC2022}, Chapman et al., in prep.). The sources span stellar mass $M_\star=(5-70)\times10^{10}$~M$_\odot$ and SFR = 400 -- 1700 M$_\odot$ yr$^{-1}$, with a median SFR of 1010$^{+560}_{-330}$~M$_\odot$~yr$^{-1}$ \citep{FriasCastillo2023,Liao2024}. {$M_\star$ and SFR were derived by \citet{Dudzeviciute2020, simpson2020} using the high-redshift version of the \textsc{MagPhys} spectral energy distribution modelling code \citep{daCunha2015, Battisti2019}.} Most of them are on the massive end of the main sequence of star-forming galaxies, but several are offset to the starburst regime (SFR/$M_\star\approx4\times$ than the main-sequence, \citealt{FriasCastillo2023}). None of the galaxies is strongly gravitationally lensed, as confirmed by ALMA 0.3''-resolution continuum imaging \citep{stach2018,simpson2020}. We can thus directly measure their intrinsic sizes. 

\subsection{JVLA observations}
The AS2VLA survey used JVLA to observe the CO(1--0) emission (rest-frame frequency: $\nu_\mathrm{rest}$ = 115.2712 GHz) in thirty DSFGs at $z=2.2-4.5$. The surveys comprises of the JVLA programs ID:21A-254 (PI: Hodge) and ID:17A-251 (PI: Walter). The observations were taken between March 2021 and December 2023, under good weather conditions in the D-array configuration. The total time per source including overheads was four~hours for almost all sources. The exceptions were AS2COS14.1, which was observed for eight hours, and AS2COS44.1
and AS2COS65.1, which were observed for two hours each.

We observed in the K- and Ka-bands, using a contiguous bandwidth of 2 GHz with 2-MHz spectral channels. Depending on the source redshift, the angular resolution ranged from 2.5$"$ to 5.1$"$ (circularised FWHM); the maximum recoverable scale was 44$"$--60$"$. We manually inspected and calibrated the data using the Common Astronomy Software Applications package (\textsc{Casa}, \citealt{casa}) versions 6.4.1 and 6.4.3 and the standard JVLA pipeline. 

{The observations of the first 17 galaxies (11 robust detections) were presented in \citet[][Paper~I]{FriasCastillo2023}.  The remaining sources from the completed survey will be presented in a forthcoming paper (Paper III); out of these, an additional \textit{eight} sources were found to have secure detections of the CO(1-0) emission line. For this work (Paper II), we selected all the sources of the entire sample having secure CO(1-0) detections (peak S/N$\geq$3 in naturally-weighted cleaned maps of CO(1-0) emission), amounting to a total of \textit{19} sources.}

\subsection{Visibility-plane analysis and stacking}

We measure the source sizes directly in the visibility-plane. This approach bypasses several issues with the \textsc{Clean}ed images, including varying beam shape, the fact that cleaned images combine the dirty and clean (Gaussian) beam, and the correlated noise in the image-plane. For illustration, we show the stacked moment-0 image in Fig.~\ref{fig:stack_image}. {We achieve an rms sensitivity of $\sigma=$0.014 Jy~km~s$^{-1}$ beam$^{-1}$ with a median beam FWHM 4.2''$\times$2.7'', corresponding to $\Sigma_\mathrm{CO(1-0)}\simeq1.1\times10^7$ K~km~s$^{-1}$~pc$^2$~kpc$^{-2}$ at $z=3.1$, the median redshift of our sample.} Image-plane deconvolution with the \texttt{imfit} task in \textsc{Casa} does not yield a reliable source size.

For the visibility-plane analysis, we process the data as follows: (1) if the wide-band imaging shows a significant (S/N$\geq$5) continuum, subtract the continuum (constant signal) from the data using the \texttt{uvcontsub} task; (2) select the velocity range within $\pm$1 FWHM from the systemic velocity (using linewidths from mid-$J$ CO lines, which are detected at a higher S/N); (3) frequency-average the selected data into a single channel; (4) bin the resulting visibilities in radial bins with a size of 2.5, 5, and 10 k$\lambda$. The fields were centred at the ALMA dust continuum peak (accurate to 0.1$"$). The phase-tracking centre is generally well-aligned with the CO(1--0) emission; we phase-shifted three sources with noticeable offsets (AEG2, CDFN1, CDFN2). For two sources (AS2COS14.1, AS2COS31.1), we found a bright continuum source within the VLA primary beam. We subtracted these companions from the visibilities in \textsc{Casa}, using the \texttt{ft} task. 

We fit each binned visibility function with: (1) point-source (constant signal) and (2) an exponential profile.  We prefer the exponential profile to the more widely-used Gaussian profile as it provides a more physical description of gas surface brightness distribution in star-forming galaxies \footnote{For the literature source sizes derived using Gaussian profiles, we convert the half-light radius from the Gaussian to exponential profile.}
(e.g., \citealt{Leroy2009, WangLilly2022}) as well as dust and stellar light distribution (e.g., \citealt{hodge2016, gullberg2019, Gillman2024}. 
We utilise Monte-Carlo Markov chain modelling using the \texttt{emcee} package \citep{emcee2013}. We  assumed wide, uniform priors in total flux and half-light radius. For each galaxy, we calculate the Akaike and Bayesian information criteria (AIC, BIC) for each model.

Table~\ref{tab:table_sizes} lists the source sizes inferred from the MCMC modelling. For galaxies for which the extended source is not preferred by both AIC and BIC, we give upper limit on the source sized, based on the 95th percentile from the posterior of the exponential model.

\begin{table*}
    \centering
 
    \caption{Spectroscopic redshifts and CO(1--0) and [\ion{C}{i}](1--0) half-light radii for individual galaxies from our sample, derived from the $uv$-plane modelling. The sizes are derived using a 5-k$\lambda$ radial bin size, and integrated over $\pm$1~FWHM bandwidth (we adopt mid-$J$ CO linewidths from \citealt{birkin2021, AS2COSPEC2022}). For galaxies for which a point-source solution is preferred, the upper limit is based on the 95th percentile from the posterior of the exponential model. {At the bottom of the table, we include two sources that have not been detected in the CO(1--0) emission, but which have [\ion{C}{i}](1--0) detections}.}
    \begin{tabular}{l|c|cccc}
        \hline
        \hline
        Source ID & $z_\mathrm{spec}$ & \multicolumn{2}{c}{CO(1--0)} & \multicolumn{2}{c}{[\ion{C}{i}](1--0)} \\
        & & [arcsec] & [kpc] & [arcsec] & [kpc]\\
        \hline
        AS2COS1.1 & 4.625 & $\leq$2.62 & $\leq$17 & $\leq$1.36 & $\leq$9\\
        AS2COS8.1 & 3.581 & $\leq$1.05 & $\leq$8 & $\leq$0.75 & $\leq$6\\
        AS2COS13.1 & 2.608 & 0.41$^{+0.16}_{-0.19}$ & 2.9$^{+1.2}_{-1.4}$ & --- &  ---\\
        AS2COS14.1 & 2.921 & 0.74$^{+0.25}_{-0.21}$ & 5.8$^{+2.0}_{-1.7}$ & 0.98$^{+0.35}_{-0.31}$ & 7.7$^{+2.8}_{-2.4}$\\
        AS2COS23.1 & 4.341 & $\leq$1.89 & $\leq$13 & 1.11$\pm$0.44 & 7.6$\pm$3.0\\
        AS2COS28.1 & 3.097 & 0.66$^{+0.28}_{-0.27}$ & 5.1$^{+2.2}_{-2.1}$ & --- &  ---\\
        AS2COS31.1 & 3.643 & $\leq$5.26 & $\leq$39 &  $\leq$0.44 & $\leq$3.4 \\
        AS2COS54.1 & 3.174 & $\leq$1.60 &  $\leq$12 & --- & ---\\
        AS2COS65.1 & 2.414 & $\leq$0.68 & $\leq$5.6 & 0.38$\pm$0.11 & 3.1$\pm$0.9\\
        AS2COS66.1$^a$ & 3.247 & $\leq$2.09 & $\leq$17 & --- & ---\\
        AS2COS139.1 & 2.219 & $\leq$1.08 &  $\leq$8 & --- & ---\\
        AS2UDS10.0 & 3.169 & $\leq$3.65 & $\leq$28 & --- & ---\\
        AS2UDS11.0 & 4.073 & $\leq$2.37 & $\leq$38 & --- & ---\\
        AS2UDS12.0 & 2.520 & $\leq$0.80 & $\leq$6.6 & 0.40$\pm$0.19 & 3.3$\pm$1.6\\
        AS2UDS126.0 & 2.436 & 0.97$^{+0.37}_{-0.25}$ &8.0$^{+3.1}_{-2.1}$ & 0.64$\pm$0.28 & 5.3$\pm$2.3\\
        AEG2 & 3.668 & $\leq$1.43 & $\leq$12 & --- & ---\\
        CDFN1 & 3.149 & $\leq$0.80 & $\leq$6 & --- & ---\\
        CDFN2 & 4.422 & $\leq$1.35 & $\leq$9 & --- & ---\\
       {CDFN8} & 4.144 & $\leq$0.89 & $\leq$6 & --- & ---\\
       \hline
    {AS2COS2.1} & 4.595 & --- & --- & 2.30$^{+1.90}_{-1.25}$ & 15$^{+12}_{-8}$\\
    {AS2UDS.26} & 3.296 & --- & --- & 0.67$^{+0.31}_{-0.30}$ & 5.1$^{+2.4}_{-2.3}$ \\
         \hline
         \hline
         \multicolumn{6}{l}{$^a$ The [\ion{C}{i}] line in AS2COS66.1 falls partially outside the ALMA tuning} \\
         \multicolumn{6}{l}{and is not included in this analysis.}
    \end{tabular}
    \label{tab:table_sizes}
\end{table*}

\section{Analysis and results}

Fits to individual galaxies show that the point-source model (our null hypothesis) is preferred for 15 out of 19 sources. {For four sources, exponential profiles with $R_\mathrm{1/2}=0.4-1.0$$"$ are preferred (see Table~\ref{tab:table_sizes}): these are AS2COS13.1, AS2COS14.1, AS2COS28.1\footnote{AS2COS28.1 has a companion offset by $\approx$20~kpc, which might contribute to the large apparent source size \citep{Chen2022}.}, and AS2UDS126.0.} These four sources do not stand out from the rest of the sample in redshift, SFR or $M_\mathrm{mol}$. We hypothesise that our detections of extended emission in these source are due to their lower-than-average redshifts compared to the whole sample (increasing the line signal), which also place the CO(1--0) line at frequencies where JVLA has particularly low noise. In particular, AS2COS13.1 and AS2COS14.1 have the highest S/N out of the entire sample.

We stack the 19 galaxies with CO(1--0) detections and fit them with (1) a point-source, (2) an exponential profile and (3) a combination of an exponential profile + a point-source (to capture a potential combination of a compact and extended component). The exponential model is preferred by both AIC ($\Delta$AIC=27) and BIC ($\Delta$BIC = 27) over the point-source model. The combination of an exponential and a point-source model is not preferred over a single exponential profile.

Changing the bandwidth used for calculation from 2 FWHM to 1 FWHM results in $\leq$10\% change in the inferred $R_\mathrm{1/2}$. Similarly, removing baseline longer than 65~k$\lambda$ -- which might have less robust calibration -- results in $\leq$10\% change. Our results are thus robust with respect to the inclusion of the longest baselines. As our fiducial value, we adopt $R_{1/2}=0.49\pm0.07$$"$ (3.8$\pm$0.5~kpc at $z=3.1$), derived for a 2$\times$FWHM bandwidth using all baselines (Fig.~\ref{fig:stack_image}, right).

Is the inferred mean source size driven by the four extended sources? As a consistency check, we perform the stacking procedure excluding the four individually extended sources: the preferred model is a single exponential profile with $R_{1/2}=$ 0.50$\pm$0.10$"$, i.e., almost identical to our fiducial model.

Finally, we stress that the extent of our sources is not due to uncertainties in position and VLA pointing accuracy. The source positions were pin-pointed by sub-arsecond ($\approx$0.3$"$) ALMA imaging; the VLA positional accuracy is typically 10\% of the beam FWHM, i.e. $\leq$0.35$"$ for an average observation from our sample. 
Our tests show that even an extreme 0.5$"$ pointing error yields a stacked source size of $\approx$0.09$"$, much smaller than the one inferred from our data.

\subsection{Spectral stacking: no signatures for CO(1--0) inflows or outflows}
\label{subsec:spectral_stack}
Are the extended CO reservoirs associated with molecular outflows or inflows? Gas outflows -- indicated by high-velocity wings of the line profile -- have been invoked to explain extended [\ion{C}{ii}] reservoirs around high-redshift galaxies from the ALPINE sample \citep{ginolfi2020, pizzati2020, Pizzati2023}. Although ALPINE galaxies have SFRs an order-of-magnitude lower than our sample (10--100~M$_\odot$ yr$^{-1}$), the same mechanism might be at play at even more highly star-forming DSFGs.

\begin{figure}[h]
    \centering
    \includegraphics[width=0.48\textwidth]{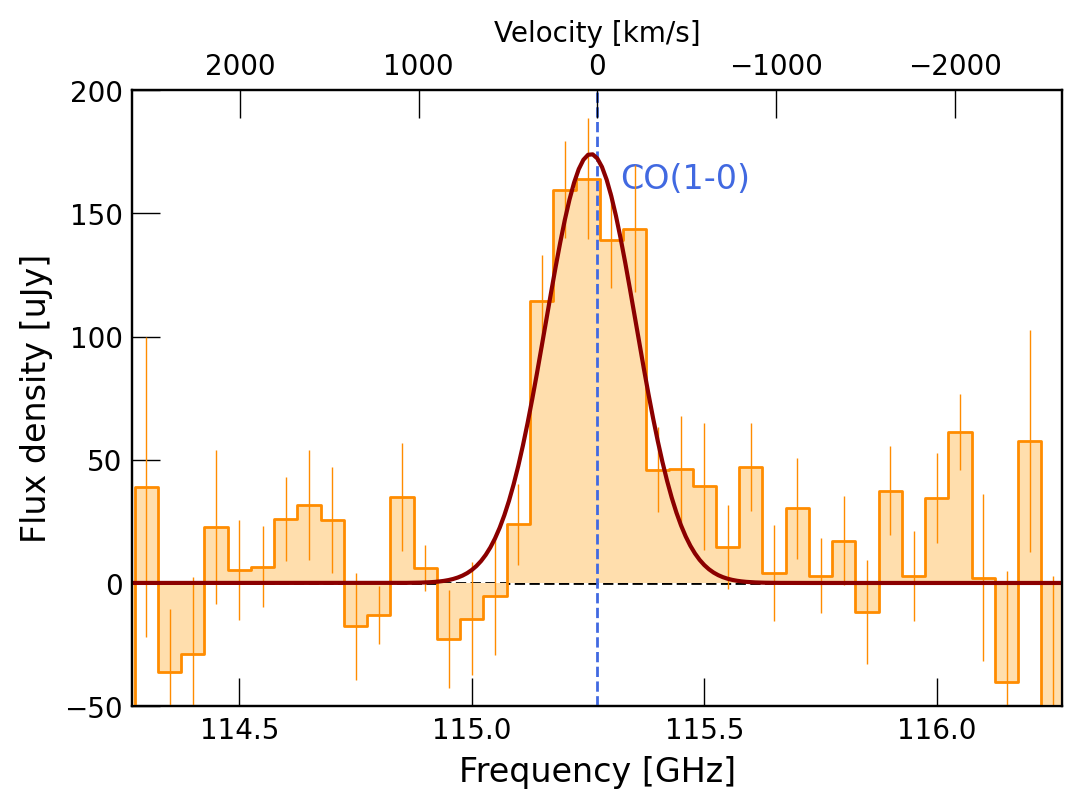}
    \caption{Rest-frame stacked spectrum, extracted from Cleaned cubes and normalised to the median flux-weighted redshift $z=3.1$. The rest-frame frequency resolution is 50 MHz ($\approx$125~km~s$^{-1}$). The stacked spectrum is consistent with a Gaussian profile with a FWHM of $590\pm70$ km~s$^{-1}$; we do not find any evidence for outflow signatures. The slight positive /negative excess at higher/lower frequencies is likely a weak residual continuum signal and is not associated with the CO(1--0) line.}
    \label{fig:spectral_stack}
\end{figure}

To search for potential high-velocity components, we stack the spectra of all 19 CO(1--0)-detected galaxies. We adopt \textsc{Clean}ed cubes, made with natural weighting. As the half-light radius of the stacked emission is smaller than the VLA beam, we extract spectra from an aperture equal to the beam size. The stacked spectrum is well-fit by a single Gaussian profile with a line FWHM of 590$\pm$70 km~s$^{-1}$ (Fig.~\ref{fig:spectral_stack}). Although our data is continuum-subtracted, we still find a low positive excess ($\approx$~5$\mu$~Jy, increasing towards higher frequencies) across the entire spectrum. This residual continuum emission is likely due to curvature of the spectrum: at the rest-frame 115~GHz, the combination of free-free and dust thermal emission can lead to a significant spectral curvature. The lack of obvious outflow features indicates that the bulk of CO(1--0) is close to the systemic velocity.

\subsection{ALMA [CI](1--0) data}
\label{subsec:alma_ci}

We supplement our CO(1--0) data with [\ion{C}{i}](1--0) ALMA observations of {ten galaxies from our parent sample} detected in [\ion{C}{i}](1--0) line by \citet[][AS2COS31.1]{Chen2022} and \citet[][nine galaxies\footnote{The tenth galaxy detected in \citet{Frias2024b} -- AS2COS66.1 -- has only part of the [\ion{C}{i}](1--0) line covered by the ALMA tuning; consequently, we do not include it in our analysis}]{Frias2024b}. The [\ion{C}{i}](1--0) line is considered an alternative tracer of molecular gas, including the ``CO-dark'' phase (e.g., at low metallicities). The \citet{Frias2024b} ALMA observations had baselines extending up to 140~k$\lambda$, giving a slightly higher angular resolution (2.2$"$-4.0$"$ circularised beam FWHM) than our JVLA data.

Fitting the [\ion{C}{i}] visibilities of individual sources, we find that five are consistent with an exponential profile, with half-light radii between 3$\pm$1 and 9$\pm$3~kpc (see Table~\ref{tab:table_sizes}). The two sources that are extended in both CO(1--0) and [\ion{C}{i}](1--0) -- AS2COS14.1 and AS2UDS126.0 -- have consistent sizes in both tracers.

{Stacking all the data, we find that the exponential profile with $R_{1/2}$=0.20$\pm$0.11$"$ (1.6$\pm$0.8~kpc) is preferred over the point-source one, albeit at low significance ($\Delta$AIC$\approx$2). The inferred mean [\ion{C}{i}](1--0) size is 60$\pm$25\% smaller than the CO(1--0) one, as a 2.3$\sigma$ significance. Given the very weak preference for the exponential profile over the point-source one, we conservatively put an upper limit on the [\ion{C}{i}](1--0) size by taking the 95th percentile of the posterior ($\leq$ 0.33$"$ (2.6 kpc). The [CI](1-0) therefore appears to be marginally less extended than CO(1-0).}

\begin{figure}[h]
    \centering
     \includegraphics[width=0.45\textwidth]{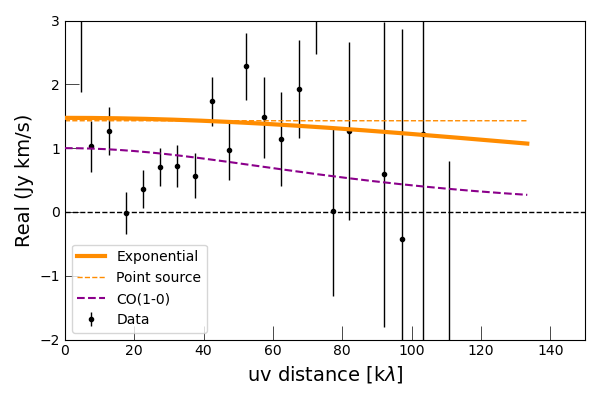}\\
        
    \caption{Visibility-plane stack of the [\ion{C}{i}](1--0) data for {ten} sources from \citet{Chen2022, Frias2024b}. The $uv$-plane data are integrated over $\pm$1~FWHM velocity range, radially binned with a step of 5~k$\lambda$. The exponential model with $R_{1/2} = 0.28\pm0.10$$"$ (2.2$\pm$0.8~kpc, solid orange line) is marginally preferred by the evidence over the point-source one (dashed). We also show the best-fitting CO(1--0) profile (magenta) for comparison.}
    \label{fig:stack_image_ci}
\end{figure}

\subsection{PDR modelling}
\label{subsec:pdr}

We now infer the thermodynamics of the molecular gas reservoirs using photo-dissociation region (PDR) modelling. Specifically, we estimate the gas density and far-UV irradiation using the \texttt{PDRToolbox} suite of uniform-density PDR models \citep{kaufman2006,pound2008}. As we will show in Section~4, the CO(1--0)/[\ion{C}{i}](1--0) sizes inferred from the uv-plane stack are consistent with the extended component of the [\ion{C}{ii}] and far-IR continuum emission seen in previous studies of DSFGs.

For the far-IR continuum, we assume that the extended component accounts for 13\% of the total luminosity (\citealt{gullberg2019}, see below). As our sample currently lacks ancillary observations in, e.g., [\ion{C}{ii}] 158-$\mu$m line, we adopt the surface brightness range reported for resolved observations of $z=2-4$ DSFGs \citep{gullberg2018, rybak2019, Mitsuhashi2021}: [\ion{C}{ii}]/FIR $=(1.5-38)\times10^{-3}$. We assume that 20\% of the [\ion{C}{ii}] emission arises from ionised gas (as seen in nearby star-forming galaxies, \citealt{Croxall2017, Sutter2019}), and apply geometric corrections to the (optically thin) predicted [\ion{C}{ii}] and FIR continuum by multiplying them by a factor of two. Finally, we include the [\ion{C}{i}](1--0) line, assuming it is co-spatial with CO(1--0).

\begin{figure}
    \centering
    \includegraphics[width=0.8\linewidth]{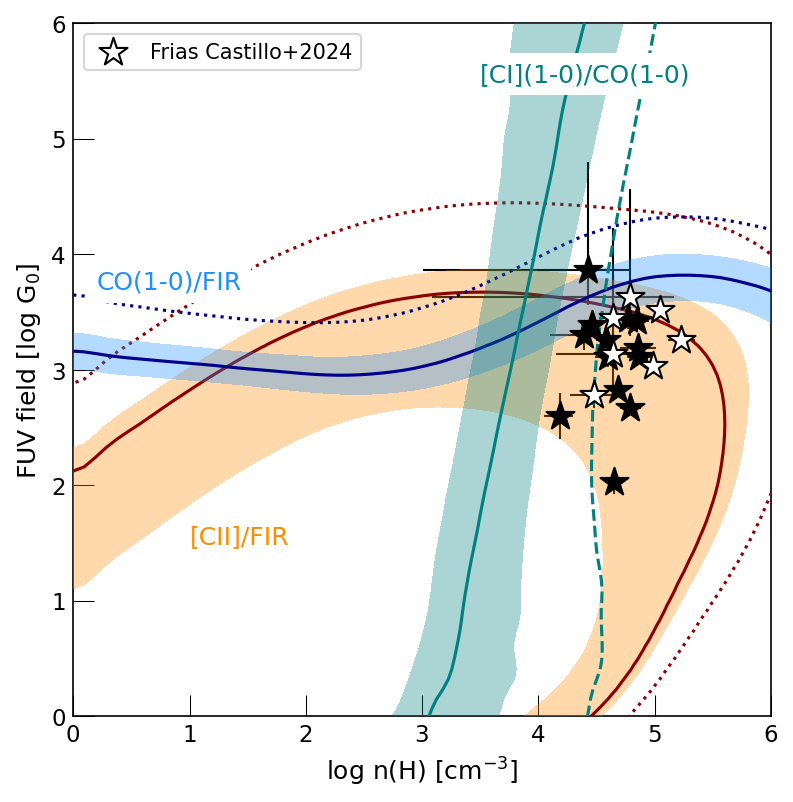}
    \caption{PDR model for the extended CO(1--0) reservoirs, compared to source-averaged models for DSFGs from the AS2VLA sample from \citet[][stars]{Frias2024b}. {The sources from \citet{Frias2024b} that are part of our sample are indicated by solid symbols; those that are not have open symbols.} Besides our CO(1--0) stacked data, we use [\ion{C}{ii}] and far-IR continuum from the literature (see main text for details). The blue dotted line corresponds to CO(1--0) being suppressed by 50\% due to elevated CMB background. The red dashed line corresponds to 20\% [\ion{C}{ii}] arising from the PDRs (as opposed to the fiducial value of 80\%). {The line and continuum ratios are consistent with an elevated density ($n=10^{4.4\pm0.4}$~cm$^{-3}$) and a high far-UV illumination ($G=10^{3.4\pm0.4}$~$G_0$), indicating they originate in dense, star-forming regions, rather than in a diffuse, quiescent gas.} }
    \label{fig:pdr}
\end{figure}

Figure~\ref{fig:pdr} shows the resulting PDR models, compared to the models for \emph{individual} galaxies from the AS2VLA sample from \citet{Frias2024b}. For the stacked data, we find a median $n=10^{4.4\pm0.4}$~cm$^{-3}$, $G=10^{3.4\pm0.4}$~$G_0$ (1~$G_0=1.6\times10^{-3}$ erg s$^{-1}$ cm$^{-2}$). These are much higher than expected from a supposedly cold, diffuse gas and comparable to values found in Galactic star-forming clouds (e.g., \citealt{Oberst2011}). The most direct interpretation is that the extended CO(1--0) (and [\ion{C}{ii}]) emission arise from in-situ star formation. This interpretation is consistent with results from resolved imaging of H$\alpha$ emission -- which traces ionised gas -- around $z\approx2$ DSFGs, which revealed extended star formation with a half-light radius of $\approx4$ kpc (e.g., \citealt{Chen2020, Birkin2024}).

To assess the sensitivity of our conclusions to the various assumptions made in the PDR modelling we have varied these as follows:

\begin{itemize}
    \item What if CO(1--0) is suppressed by CMB? Assuming a 50\% suppression (an extreme scenario at $z\approx3$, e.g., \citealt{dacunha2013}), the inferred far-UV illumination increases by $\approx$0.5~dex (blue dotted line in Fig.~\ref{fig:pdr}).
    \item What if the bulk of [\ion{C}{ii}] is arising from ionised gas rather than from the PDRs? Lowering the PDR contribution to the [\ion{C}{ii}] flux down to 20\% causes the inferred far-UV illumination to increase by almost 1~dex (red dotted line in Fig.~\ref{fig:pdr}).
     \item What if [\ion{C}{i}](1--0) is more compact than CO(1--0) (i.e., the ratio of atomic/molecular carbon increases at large radii)? We assume the [\ion{C}{i}](1--0) scale length indicated by the tentative results in Section~3.2 and re-calculate the [\ion{C}{i}](1--0) surface brightness at the CO(1--0) half-light radius accordingly. The inferred density increases to $\approx10^{4.5}$~cm$^{-3}$ (green dotted line in Fig.~\ref{fig:pdr}).
    
\end{itemize}

We derive the filling factor $\phi$ -- i.e., the number of emitting regions per line-of-sight -- by comparing the observed CO(1--0) surface brightness with the \textsc{PDRToolbox} models. Specifically, for each realisation of $G$, $n$, we calculate the corresponding filling factor at the CO(1--0) half-light radius $R_{1/2}$ as:

\begin{equation}
    \phi = \frac{L_\mathrm{CO(1-0)}}{2\pi R_{1/2}^2 \times \Sigma_\mathrm{CO(1-0)}} \times e^{-1.68} \times \frac{1}{4\pi D_L^2},
\end{equation}

where $\Sigma_\mathrm{CO(1-0)}$ is the CO(1--0) flux per solid angle predicted by \textsc{PDRToolbox} and $D_L$ is the luminosity distance. 
Marginalising over $G$ and $n_H$ then yields $\phi(R=R_{1/2})=0.8\pm0.5$\%. {We discuss the context of this result in Section~\ref{subsec:filling_factor}.}

\section{Discussion}
\label{sec:discussion}

{In this Section, we compare the results of our stacking procedure and PDR modelling to other studies of high-redshift DSFGs (\ref{subsec:sizes}), dust and stellar emission (\ref{subsec:m_mol}), and hydrodynamical simulations (\ref{subsec:sims}). Finally, we address the nature of extended CO(1--0) reservoirs (\ref{subsec:filling_factor}).}

\subsection{Sizes of extended molecular gas reservoirs in high-redshift DSFGs}
\label{subsec:sizes}

\begin{figure*}
    \centering
    \includegraphics[height=7cm]{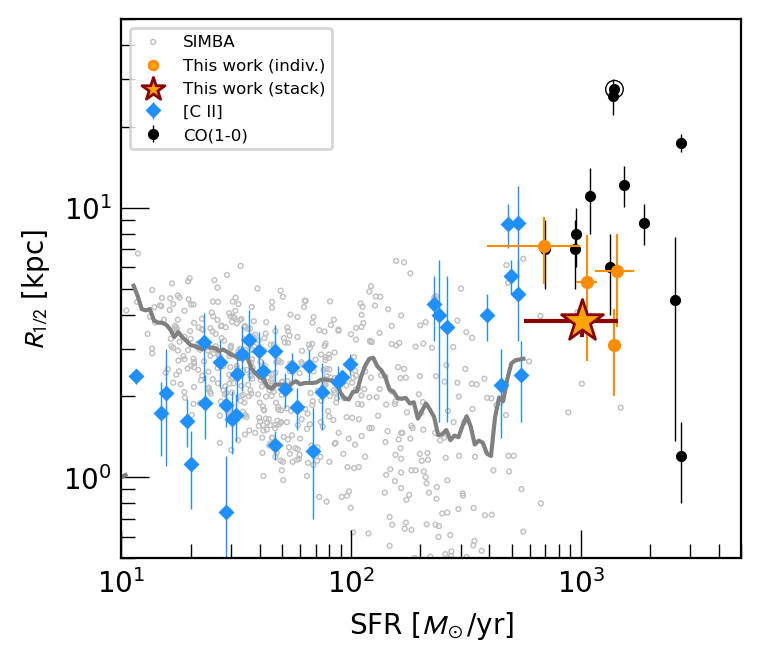}
    \includegraphics[height=7cm]{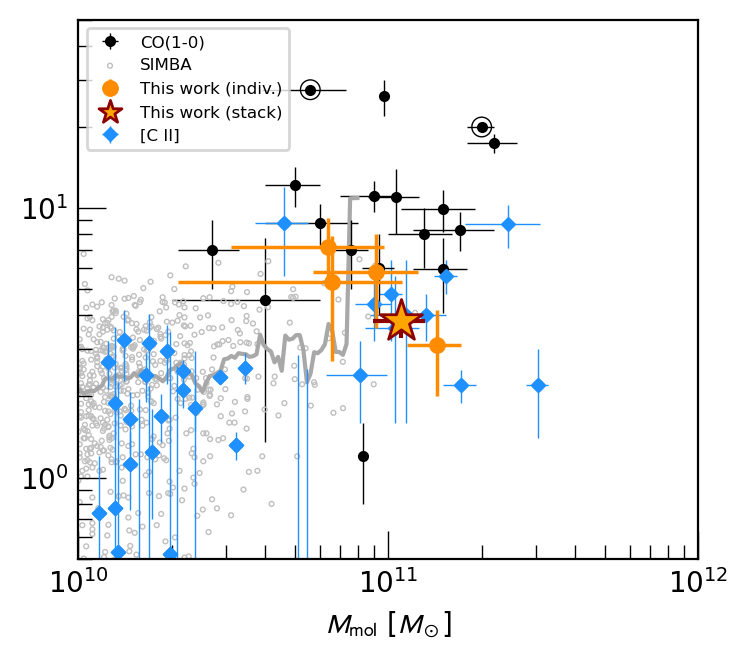}
    \caption{Half-light radii of CO(1--0) and [\ion{C}{ii}] emission versus star-formation rate (\textit{left}) and total molecular gas mass (\textit{right}) for galaxies from our sample and literature (see Section~\ref{subsec:sizes} for references). The protoclusters from \citet{emonts2016} and \citet{dannerbauer2017} are highlighted by circles; open symbols indicate lensed sources from \citet{Stanley2023}. We also show predictions for $z=3.1$ galaxies from the SIMBA simulation \citep{dave2019} as open grey symbols; {grey lines indicate the running average}. The inferred size of the CO(1--0) emission in our sample is consistent to the extended CO(1--0) and [\ion{C}{ii}] reservoirs around other high-redshift DSFGs with comparable SFR.}
    \label{fig:r_mmol_sigma}
\end{figure*}

Figure~\ref{fig:r_mmol_sigma} places our measurement of CO(1--0) sizes in context of other high-redshift observations. Specifically, we show the literature CO(1--0) \citep{ivison2011, thompson2012, dannerbauer2017, friascastillo2022,Stanley2023} and [\ion{C}{ii}] measurements \citep{gullberg2018, rybak2019, Mitsuhashi2021}. {We derive $M_\mathrm{mol}=\alpha_\mathrm{CO}\times L'_\mathrm{CO(1-0)}$ with $\alpha_\mathrm{CO}=1.0$ (e.g., \citealt{calistro-rivera2018}). For the [\ion{C}{ii}]-based masses, we use $M_\mathrm{mol}=\alpha_\mathrm{[CII]}\times L_\mathrm{[CII]}$ , assuming $\alpha_\mathrm{[CII]}$=30$M_\odot/ L_\odot$ \citep{Zanella2018}.}

Our inferred CO(1--0) size is comparable to CO(1--0) sizes reported in the literature for other highly star-forming galaxy populations, and the [\ion{C}{ii}] reservoirs seen around high-redshift DSFGs. {Our inferred sizes are also consistent with the CO(1--0) sizes reported by \citet{Stanley2023} for the (potentially) non-lensed sources from the Z-GAL sample (median $R_{1/2}=$4.4~kpc). In contrast, our CO(1--0) size is larger than the typical extent of CO(3--2) emission in $z=2-3$ DSFGs---$R_\mathrm{1/2}\leq$2.6~kpc---recently reported by \citet{Amvrosiadis2025}.} The true extent of the molecular gas reservoirs in DSFGs might be even larger than those inferred from our CO(1--0) stacking. First, gas becomes less metal-rich with increasing distance from the centre of the galaxy (e.g., \citealt{Schimek2024}), causing $\alpha_\mathrm{CO}$ to increase towards the outskirts. Moreover, the elevated CMB temperature at high redshift can significantly suppress the CO(1--0) emission, particularly at low gas kinetic temperatures and densities (by up to 50\%, e.g., \citealt{dacunha2013, zhang2016}). However, the CMB effect will be more limited if the molecular gas is mostly in the form of higher-density ISM in star-forming regions or satellite galaxies.

{The CO(1--0) extent derived from our stacking is comparable to the sizes of [\ion{C}{ii}] emission in high-redshift galaxies with similar SFR and molecular gas mass (Fig.~\ref{fig:r_mmol_sigma}, e.g., \citealt{gullberg2018, rybak2019,Mitsuhashi2021}), as well as those reported around $z=4-6$ star-forming galaxies from the CRISTAL sample ($R_{1/2}=3.2\pm1.3$~kpc, \citealt{Ikeda2025}).}

{The extended [\ion{C}{ii}] reservoirs around high-redshift galaxies have been hypothesised to arise primarily from atomic or ionised gas (e.g., \citealt{Ikeda2025}), perhaps deposited by outflows (e.g., \citealt{pizzati2020, Pizzati2023}. However, the comparable extent of the CO(1--0) (which is associated only with molecular gas) indicates that molecular gas might be the dominant component of the extended gas halos.}

\subsection{Molecular gas versus dust and stellar emission}
\label{subsec:m_mol}

How does the size of the extended CO(1--0) emission compare to the far-infrared and near-infrared continuum in DSFGs? In more physical terms, we now compare the molecular gas sizes to star formation (traced by far-IR continuum) and stellar emission (traced by JWST NIRCam imaging). Table~\ref{tab:sizes_co_dust_stars} lists the sizes of different tracers.

Using high-resolution 870-$\mu$m ALMA imaging of the SCUBA-2 sources in the UKIDSS/UDS field (from which our sample is partially drawn), \citet{gullberg2019} found a half-light radius of 1.1$\pm$0.3~kpc, assuming an exponential profile\footnote{After applying a correction to account for a faint extended component \citep{Smail2021}.}. Somewhat larger sizes (1.8$\pm$0.1~kpc) were inferred for the ALESS sample of DSFGs \citep{hodge2016, Hodge2024}, which span a comparable redshift range (median $z= 3.4_{-1.0}^{+0.4}$) and 850-$\mu$m flux ($S_\mathrm{850\,\mu m} =6.4_{-3.0}^{+2.0}$~mJy) to our galaxies. In either case, the 870-$\mu$m continuum is a factor of $\approx 2-3$ more compact than the CO(1--0) reservoir. 

In total, 50--80\% of the CO(1--0) emission arises \emph{outside} the region which contains 90\% of the dust continuum emission (adopting the \citealt{gullberg2019} and \citealt{hodge2016} sizes, respectively). 
The remarkable difference in sizes between the dust continuum and CO(1--0) emission indicates that only a small fraction of the total molecular gas in high-redshift DSFGs contributes directly to their high SFRs. This interpretation is supported by recent studies of HCN/HCO$^+$/HNC emission in $z\geq2$ DSGFs, which found very low dense-gas fractions \citep{Rybak2022}.

The presence of an extended molecular gas dovetails with the extended low-surface brightness dust continuum reported by \citet{gullberg2019}, {who after stacking ALMA imaging of 153 DSFGs, reported that $\approx$13\% of the 870-$\mu$m continuum arises from an extended component with an assumed radius of $\approx$4 kpc, which is similar to our measured CO(1--0) size.} This extended continuum emission lends further support to the presence of significant metal-enriched gas and reservoirs around high-redshift DSFGs.

As the bulk of the molecular gas is far outside the central star-forming region, can the high SFRs of DSFGs be sustained by smooth gas accretion from this reservoirs? We can estimate the required inward streaming velocity by dividing the half-light radius by 
{one half of the} median gas depletion time ($t_\mathrm{dep}=140\pm70$~Myr, \citealt{FriasCastillo2023}). We obtain $v_\mathrm{in}=50\pm12$ km\,s$^{-1}$, consistent with to the line-of-sight velocity dispersions seen in high-resolution kinematic studies of $z=2-4$ DSFGs (10 - 100 km\, s$^{-1}$, e.g., \citealt{hodge2012, rybak2015b, swinbank2015, rizzo2021, rizzo2023, Birkin2024, Amvrosiadis2025}. The central star formation can thus be supported by smooth streaming of gas within the galaxy without major mergers (see, e.g., \citealt{narayanan2015}).

For the sizes of stellar emission, we adopt the 4.4-$\mu$m half-light radius of $R_{1/2}=2.7\pm0.2$~kpc derived by \citet{Gillman2024} for the AS2COSMOS and AS2UDS DSFGs.
A similar value (3.0$\pm$0.3~kpc) was found for the ALESS sample by \citet{Hodge2024}. The CO(1--0) emission in our sample is a factor of 1.4$\pm$0.2 and 1.3$\pm$0.2 more extended than the 4.4-$\mu$m sizes from \citet{Gillman2024} and \citet{Hodge2024}, respectively. This size discrepancy indicates that in DSFGs, the cold molecular gas appears to extend beyond the stellar emission of the galaxy.

\begin{table}[]
    \caption{Comparison of CO(1--0) sizes inferred in this work, compared to [\ion{C}{ii}], ALMA 870-$\mu$m continuum and JWST 4.4-$\mu$m continuum. We list the 870-$\mu$m and 4.4-$\mu$m sizes for AS2UDS ans AS2COSMOS surveys \citep{gullberg2019, Gillman2024} -- the parent samples of the AS2VLA survey -- and the ALESS survey \citep{hodge2016, Hodge2024}.}
    \label{tab:sizes_co_dust_stars}
    \centering
    \begin{tabular}{c|cc}
    \hline
         Tracer & $R_{1/2}$ [kpc] & Reference \\
    \hline
        CO(1--0) & 3.8$\pm$0.5 & This work \\
        {[\ion{C}{i}](1--0)} & $\leq${2.6} & This work \\
        870 $\mu$m & 1.1$\pm$0.3 & \citet{gullberg2019} \\
        870 $\mu$m & 1.8$\pm$0.1 & \citet{hodge2016} \\
        4.4 $\mu$m & 2.7$\pm$0.2 & \citet{Gillman2024}\\
        4.4 $\mu$m & 3.0$\pm$0.3 & \citet{Hodge2024}\\
    \hline
    \end{tabular}
    
\end{table}

The discrepancy in the spatial extent of molecular gas versus star formation and stars might be even more pronounced than indicated from the CO(1--0) and 4.4-$\mu$m imaging. First, molecular gas might be more extended than CO(1--0) emission, due to higher $\alpha_\mathrm{CO}$ and increased CMB attenuation in the low-density outskirts of galaxies. Second, if the dust temperature decreases towards larger radii (e.g., \citealt{calistro-rivera2018}) star formation will be more centrally concentrated than the observed-frame 870-$\mu$m continuum.
Finally, the stellar component is likely more compact than the near-infrared continuum; high obscuration in the central region will ``inflate'' the apparent near-infrared sizes compared to the actual stellar component \citep{Sorba2018, popping2022, Smail2023, Gillman2024}.

How do the relative sizes of molecular gas, stars and dust continuum in high-redshift DSFGs compare to other galaxy populations? At $z\sim0$, spiral galaxies have molecular gas reservoirs with half-light radii of $\approx$1--5~kpc (e.g., \citealt{Regan2001,Leroy2009}). The CO(1--0)/(2--1) emission in spiral galaxies is typically smaller than the stellar emission (e.g., \citealt{Regan2001} find a mean ratio of 0.88$\pm$0.14). In contrast, the $z\approx0$ ultraluminous infrared galaxies (ULIRGs) have very compact molecular gas reservoirs (typically less $\leq$1~kpc, e.g., \citealt{Bellocchi2022}), a factor of a few smaller than their stellar components. The larger extent of molecular gas compared to the stellar component in DSFGs suggests an inside-out growth (as implied by the small dust continuum sizes) or a presence of satellite galaxies.

\subsection{Comparison to molecular gas in hydrodynamical simulations}
\label{subsec:sims}

How do our findings compare to predictions from simulations? Diffuse molecular gas has been notoriously difficult to produce in large-volume hydrodynamical simulations, as convergence requires very high (pc-scale) resolution (e.g., \citealt{FaucherOh2023}).

For a comparison in Fig.~\ref{fig:r_mmol_sigma}, we take data from the SIMBA cosmological simulation \citep{dave2019}. SIMBA successfully reproduces key parameters of high-redshift DSFGs \citep{Lovell2021}. We focus on the $z=3.1$ snapshot from the default 100-cMpc box. We select galaxies with $M_\mathrm{mol}\geq 1\times10^{11}$~M$_\odot$, i.e., comparable in mass to our sample. Specifically, we consider the molecular gas masses and the (total) gas half-light radii from the SIMBA catalogue. We note that SIMBA does not predict H$_2$ directly; rather, the molecular gas is predicted by post-processing the simulation data \citep{dave2019}. The gas sizes are determined by averaging two-dimensional projections along the three axes. 

As shown in Fig.~\ref{fig:r_mmol_sigma}, SIMBA galaxies with the highest molecular-gas masses have sizes and mean surface densities comparable to our sample. {However, unlike the observed galaxies, SIMBA galaxies tend to become more compact with increasing SFR. This discrepancy is likely a result of the sub-grid star-formation prescription used in SIMBA, which implements a Kennicutt-Schmidt law (i.e., star-formation rate scales with the gas surface density, \citealt{dave2019}).}

\subsection{The nature of the extended CO(1--0) reservoirs}
\label{subsec:filling_factor}

What is the nature of the extended CO(1--0) emission around DSFGs? Are they smooth extensions of the galaxy's gas reservoirs, tidal features induced by galaxy mergers, or individual faint satellites?
Discriminating between different scenarios requires deep, sub-arcsecond observations of individual galaxies, while preserving sensitivity to large spatial scales. Unfortunately, our stacking averages the data in the azimuthal direction - i.e., smearing emission from compact satellites.  

However, we can get significant insights using results from the PDR modelling. As shown in Section~\ref{subsec:pdr}, the inferred gas density and far-UV illumination are comparable to those in Galactic star-forming clouds. 
Conversely, the filling factor ($\phi=0.8\pm0.5$\%) is smaller than the disks of nearby galaxies, in which it ranges between few per cent to factor unity (e.g., \citealt{Wolfire1990, Walsh2002, Luhman2003}). 

The combination of high $G$, $n$ and low $\phi$ indicates that molecular gas resides in compact, dense clouds, that are distributed sparsely on the outskirts of galaxies.
Alternatively, the clouds might be concentrated in satellites with small angular separation from the host (although we note the lack of companions at $\leq$1$"$ separation in JWST NIRCam imaging of DSFGs in \citealt{Gillman2024, Hodge2024}). In either scenario, the molecular gas on the galaxy outskirts would be significantly clumpy rather than distributed in a smooth, diffuse ``halo''.

\section{Conclusions}
\label{sec:conclusions}

We have presented results of stacking CO(1--0) observations of a sample of 19 high-redshift DSFGs. Comprising $\approx$80~hours in the most compact JVLA array configuration, this is the deepest study of extended cold-gas reservoirs around early galaxies to date. Our main findings are:

\begin{itemize}
    \item The stacked $uv$-plane data of the CO(1--0) emission show an extended exponential profile with a half-light radius of 3.8$\pm$0.5~kpc. The inferred size is comparable to extended CO(1--0) and [\ion{C}{ii}] reservoirs previously reported around high-redshift DSFGs.
    \item Four galaxies have clearly extended CO(1--0) emission, with half-light radii between 3.4--7.2~kpc.
    \item {Applying the same procedure to ALMA [\ion{C}{i}](1--0) observations, we find that [\ion{C}{i}](1--0) appears more compact ($R_{1/2}\leq$2.6~kpc) than CO(1--0), although both sizes are consistent within 2$\sigma$. Five galaxies show extended [\ion{C}{i}](1--0) emission, with half-light radii of 3.1--7.7~kpc.}
    \item Using PDR modelling to infer physical conditions in the extended CO(1--0) reservoirs, we find a density of $\approx10^{4.5}$~cm$^{-3}$ and a far-UV field of $\approx10^{3.5}$~$G_0$, and a filling factor of $\approx$1\%. These results indicate that the extended CO (and [\ion{C}{i}] and [\ion{C}{ii}]) emission arises from sparsely distributed star-forming clouds.
    \item The inferred sizes of CO(1--0) cold-gas reservoir are significantly larger than the bulk of the 870-$\mu$m dust continuum, but similar to the fainter, extended dust component that has been reported in these sources. The CO(1--0) is somewhat larger than the 4.4-$\mu$m stellar emission. The bulk of the molecular gas (50--80\%) resides outside the FIR-bright star-forming region and does not directly contribute to the observed star formation.
\end{itemize}

{Our findings reveal that the bulk of molecular gas in high-redshift DSFGs resides in extended, 10-kpc scale reservoirs; only a small fraction of this gas directly contributes to the observed star formation.}



\begin{acknowledgements}
M.R. is supported by the NWO Veni project "Under the lens" (VI.Veni.202.225). J.A.H. acknowledges support from the ERC Consolidator Grant 101088676 (VOYAJ). I.R.S. and A.M.S. acknowledge STFC grant ST/X001075/1. E.F.-J.A. acknowledges support from UNAM-PAPIIT project IA102023, and from CONAHCyT Ciencia de Frontera project ID: CF-2023-I-506. C.-C.C. acknowledges support from the National Science and Technology Council of Taiwan (111-2112M-001-045-MY3), as well as Academia Sinica through the Career Development Award (AS-CDA-112-M02).

The National Radio Astronomy Observatory is a facility of the National Science Foundation operated under cooperative agreement by Associated Universities, Inc.

This paper makes use of the following ALMA data: ADS/JAO.ALMA \#2021.1.01324.S. ALMA is a partnership of ESO (representing its member states), NSF (USA) and NINS (Japan), together with NRC (Canada), MOST and ASIAA (Taiwan), and KASI (Republic of Korea), in cooperation with the Republic of Chile. The Joint ALMA Observatory is operated by ESO, AUI/NRAO and NAOJ.

The authors are thankful for the assistance from Allegro, the European ALMA Regional Center node in the Netherlands. We thank D.~Narayanan, E. Pizzati and A.~Schimek for sharing insights from theoretical models and simulations. Finally, we thank the referee -- P. Cox -- for his detailed comments that helped us improve this manuscript.\\

\end{acknowledgements}

%
\bibliographystyle{aa} 
\bibliography{mrybak_as2vla_v2} 
%

\end{document}